\documentclass[twocolumn,groupedaddress]{revtex4-2}
\pdfoutput=1

\usepackage[intlimits, sumlimits, namelimits]{amsmath}
\usepackage{amssymb} 
\usepackage{amsfonts} 
\usepackage{esdiff}
\usepackage{commath}
\usepackage{xfrac} 
\usepackage{siunitx}
\DeclareSIUnit{\molar}{M}

\usepackage{graphicx}
\usepackage{array}
\newcolumntype{M}{>{$}m<{$}} 

\begin{document}

\title{A generic self-stabilization mechanism for biomolecular adhesions under load}

\author{Andrea Braeutigam, Ahmet Nihat Simsek, Gerhard Gompper}

\affiliation{Theoretical Physics of Living Matter, Institute for Biological Information Processes,
Forschungszentrum J\"ulich, 52425 J\"ulich, Germany}

\author{Benedikt Sabass}

\affiliation{Theoretical Physics of Living Matter, Institute for Biological Information Processes,
Forschungszentrum J\"ulich, 52425 J\"ulich, Germany}
\affiliation{Institute for Infectious Diseases and Zoonoses, Department of Veterinary Sciences, Ludwig-Maximilians-Universit\"at M\"unchen, 80752 Munich, Germany}

\begin{abstract}
\noindent
Mechanical loading generally weakens adhesive structures and eventually leads to their rupture. However, 
biological systems can adapt to loads by strengthening adhesions, which is essential for maintaining the
integrity of tissue and whole organisms. Inspired by cellular focal adhesions, we suggest here a generic, 
molecular mechanism that allows adhesion systems to harness applied loads for self-stabilization under 
non-equilibrium conditions -- without any active feedback involved. The mechanism is based on 
conformation changes of adhesion molecules that are dynamically exchanged with a reservoir. Tangential
loading drives the occupation of some stretched conformation states out of equilibrium, which, for 
thermodynamic reasons, leads to association of further molecules with the adhesion cluster. 
Self-stabilization robustly increases adhesion lifetimes in broad parameter 
ranges. Unlike for catch-bonds, bond dissociation rates do not decrease with force. The self-stabilization 
principle can be realized in many ways in complex adhesion-state networks; we show how it naturally occurs 
in cellular adhesions involving the adaptor proteins talin and vinculin.
\end{abstract}

\maketitle

\section{Introduction}
\begin{figure*}
\centering
\includegraphics[width=0.98\linewidth]{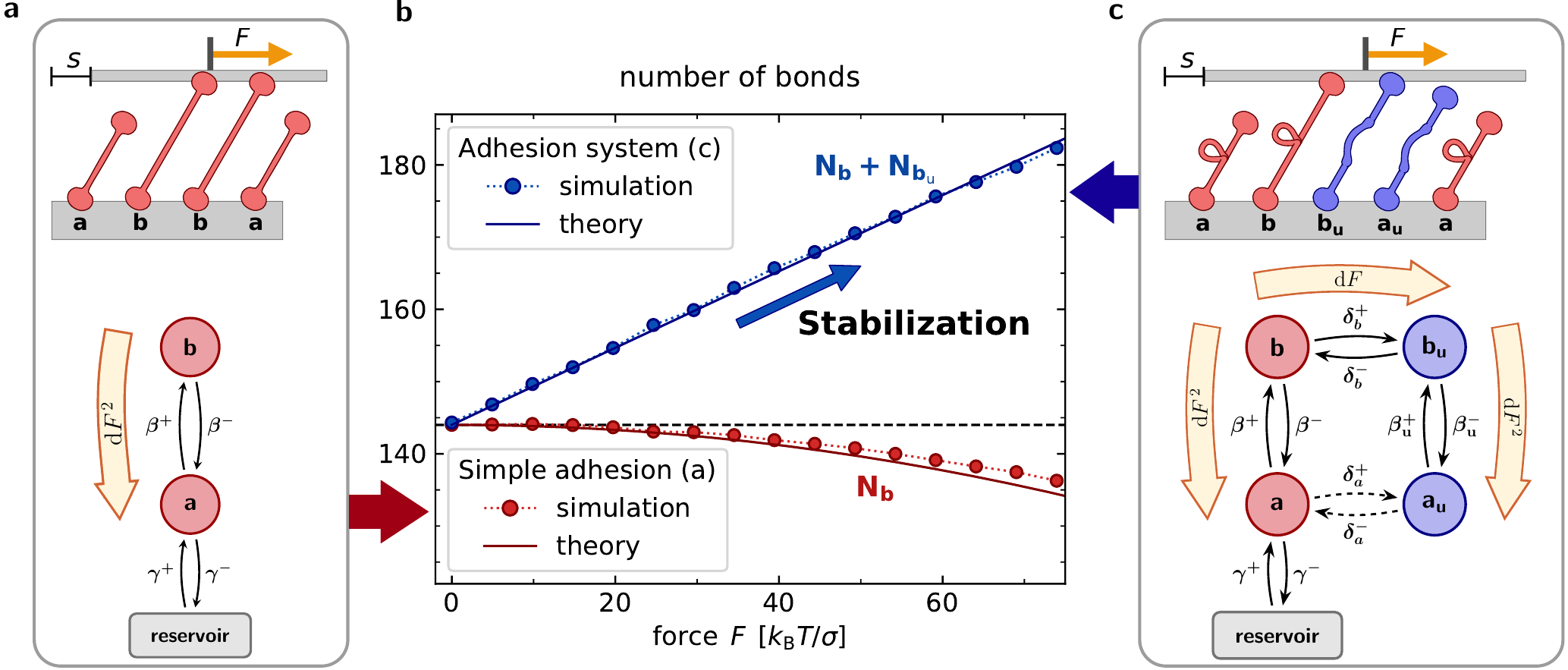}
\caption{Adhesion self-stabilization.
a)~Illustration of the basic adhesion model which consists of unbound molecules $a$ and bonds $b$ that connect two parallel rigid planes. A tangential load stretches all molecules in state $b$ as it shifts the upper boundary by $s$. Molecules can transition between states $b$ and $a$ with extension-dependent rupture and binding rates given by $\beta^-(h)$ and $\beta^+(h)$. Molecules can enter or leave the adhesion cluster with rate constants given by $\gamma^{\pm}$. b)~Mean number of bonds in steady state as a function of loading force. Symbols show simulation results and lines correspond to approximate analytical results. In the basic model, an increased load on the adhesion reduces the number of bonds (red). In contrast, an increased load produces a growth of the number of bonds in the generalized model (blue). c)~Illustration of the generalized adhesion model incorporating molecule unfolding and refolding with rates $\delta_{a,b}^\pm(h)$ as well as a molecule-exchange with the reservoir. Mechanical load drives the system out of equilibrium, shifting the state occupations. See also SI~movies~1,2.
\label{fig:intro}} 
\end{figure*}
\noindent
Multicellular organisms are held together by complex biomolecular adhesion structures. For decades, cellular adhesions have been extensively studied, both, as a paradigm for fundamental biophysical mechanisms, as well as to understand their essential biological function. Yet, a very fundamental property of cell-matrix adhesions remains mysterious – their ability to adapt their size to the mechanical load. How is it possible – from a physics perspective – that the strength of adhesions increases under load?

Cell-matrix adhesions, also called focal adhesions, are crucial for cell physiology~\cite{Geiger2009,Schoen2013}, cell motility~\cite{Huttenlocher2011}, cancer metastasis~\cite{butcher2009tense,Schwager2019}, and development~\cite{engler2006matrix,heisenberg2013forces}. The structures consist of transmembrane integrins and adaptor proteins that connect the force-generating actomyosin cytoskeleton with the extracellular matrix. Accordingly, focal adhesions have been likened to a ``molecular clutch''~\cite{Mitchison1988,gardel2008traction,moore2010stretchy, Sun2016, Chan2008,Elosegui-Artola2018}. In a pioneering experiment by Riveline et al.~\cite{riveline2001focal}, it was shown that local application of centripetal forces to adherent cells induces focal adhesion 
growth~\cite{riveline2001focal}. Moreover, the size of focal adhesions is proportional to the load~\cite{balaban2001force}. The regulatory network associated with focal adhesions is complex and several biological processes have been proposed to play a role for adhesion stability. These include mechanosensitive activation 
of integrins~\cite{Sun2016},
catch-bond behavior of integrins and vinculin-actin binding~\cite{Huang2017}, non-linear mechanical response of unfolded proteins, and downstream signaling, e.g., 
mediated by the adaptor protein p130Cas \cite{sawada2006force, wang2017review}. 
During the past years, the pivotal role of the adaptor protein talin for adhesion 
maturation has also been established~\cite{case2015integration,Goult2018}. Talin 
directly transmits forces by binding with its globular head domain to integrin, while 
its rod domain links to F-actin~\cite{Liu2015}. Under stretch, conformation changes 
in talin occur, leading to an unfolding of protein domains and to the exposure of 
cryptic binding sites for vinculin~\cite{DelRio2009,Yao2016,rahikainen2017mechanical}. 
Vinculin, in turn, further recruits F-actin and thereby strengthens the linkage~\cite{Austen2015,Atherton2015,massou2020cell}. 
Moreover, other adhesion types such as adherens junctions are also capable of a load 
adaptation based on unfolding and recruitment of further 
molecules~\cite{chen2017receptor}. In spite of the considerable amount of theoretical approaches~\cite{nicolas2004cell, shemesh2005focal, Besser2006, lee2007force, Sabass2010, Li2010,leoni2017model,sens2020stick,danuser2013mathematical} 
and pioneering work combining modeling and experiment~\cite{Chan2008,bornschlogl2013filopodial,Elosegui-Artola2016}, the physical principle underlying load-adaptation of focal adhesions remains largely not understood. 

We propose here a generic mechanism through which molecular adhesions can harness mechanical load for adapting their size and stability without active feedback. A minimal model is employed, which relies on a combination of unfolding of adhesion molecules under force~\cite{Elosegui-Artola2016,Tapia-Rojo2020}, the dynamical exchange of molecules with a reservoir~\cite{nicolas2004cell,shemesh2005focal}, and possibly the recruitment of additional molecules that stabilize the unfolded conformations, all under the constraint of thermodynamical consistency. The key property of our molecular state-networks is that some states are separated from the reservoir and driven out of equilibrium. Under tangential stress, the state occupations shift, leading to a growth of the adhesion under increasing load, until the adhesion ultimately fails at very high loads. This non-equilibrium mechanism of adaptation is simple and robust. We show how the mechanism is naturally realized by the talin-vinculin system at focal adhesions and perhaps in other bioadhesions. Moreover, the proposed mechanism is general enough that it could possibly be implemented in engineered biomimetic systems.

\section{Results}
\noindent 
\textbf{Molecular adhesion model.}~
We consider a generic adhesion system consisting of $N$ molecules that form harmonic bonds 
between two planar, parallel surfaces, see Fig.~\ref{fig:intro}. The molecule extension, 
i.e., the difference between actual length and rest length, is denoted by $h$ and the 
spring constant by $\kappa$. The bottom surface is fixed in space and a constant 
tangential loading force $F$ is exerted on the top, leading to a time-dependent tangential 
shift $s$. The model is two-dimensional and forces normal to the surface are not 
considered. The adhesion-molecule number $N$ can vary by exchange with a reservoir, as 
determined by the rate constants $\gamma^{\pm}$. Individual molecules undergo stochastic 
transitions between the adhesion cluster and the reservoir, and all transition rates are chosen to fulfill detailed balance when $F=0$, i.e., in thermal 
equilibrium, see Methods~\ref{methods:section}, which avoids unphysical energy injection 
that could produce an apparent motor-like behavior. 

In a first, basic model, see Fig.~\ref{fig:intro}a, molecules from the reservoir associate 
reversibly with the adhesion system via the state $a$, in which they have not yet formed 
a bond between the upper and lower surfaces. The state where a bond has formed is 
denoted by $b$. The bond formation rate is $\beta^+(h)$, which is maximal when the 
extension of the adhesion molecule equals an optimal binding distance $|h| = 
\ell_\mathrm{b}$. For bond formation, it is assumed that the extension $h$ fluctuates 
thermally with magnitude $\sigma = \sqrt{k_\mathrm{B}T /\kappa}$, where $k_\mathrm{B}T$ is 
the thermal energy scale. For bond dissociation, we focus on slip-bond dynamics with 
rupture rates $\beta^-(h)$ that increase exponentially with bond extension, see 
Secs.~\ref{methods:section}~A-B.

In a second, generalized model, see Fig.~\ref{fig:intro}c, the molecules can undergo 
abrupt conformational changes to unfolded states, denoted by a subscript $u$, such that 
$a_\mathrm{u}$ are the unfolded, unbound states and $b_\mathrm{u}$ the unfolded, bound states. The 
overall number of bound molecules, irrespective of their conformational state, is $N_B = N_b + 
N_{b_\mathrm{u}}$, and the corresponding overall number of unbound molecules is $N_A =  N_a + 
N_{a_\mathrm{u}}$. Details of the reaction rates are given in Methods~\ref{methods:section}~A-C. 
We assume that mechanical relaxations occur instantaneously and viscous damping is neglected, so 
that the sum of the forces borne by the bonds equals the applied load $F$ at all times. 
Stochastic bond dynamics is simulated with an exact algorithm, see Supplementary Information II. 

\begin{figure*}[hbtp]
	\centering
	\includegraphics[width=\linewidth]{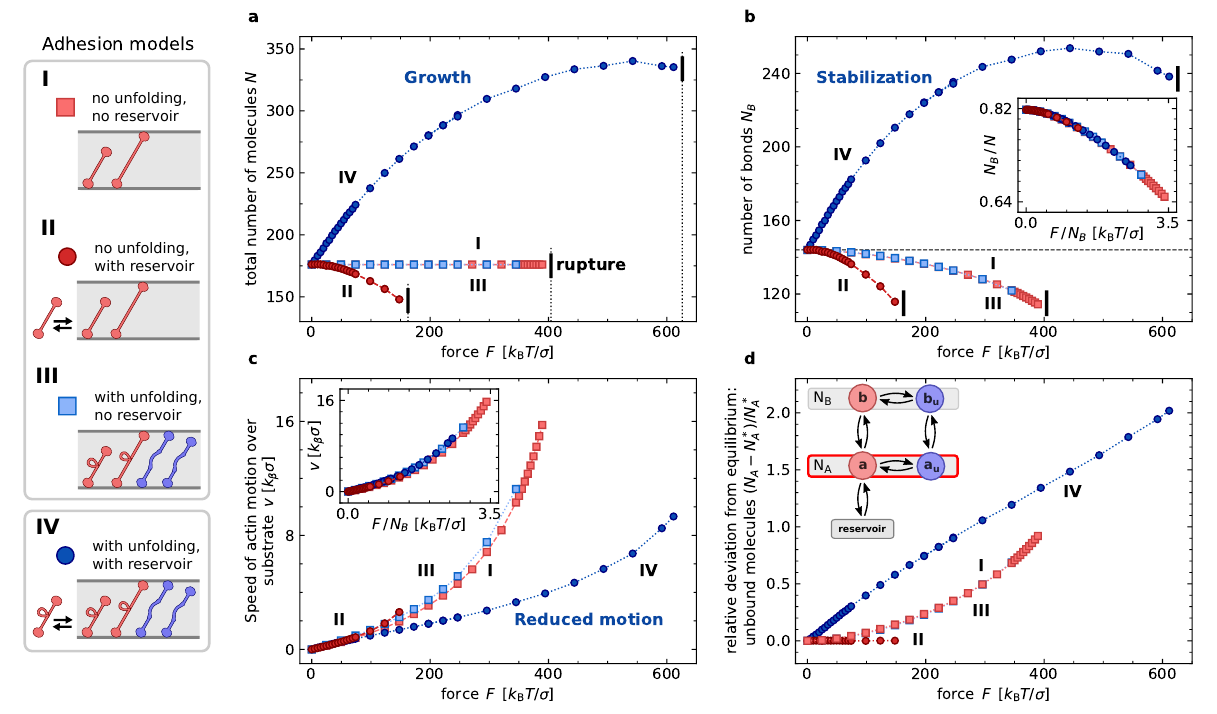}	
	\caption{
Comparison of models I-IV for adhesions with and without molecule unfolding and reservoir exchange. 
Steady-state quantities are plotted until forces at which first complete adhesion ruptures occur.
a)~Averaged total number of molecules $N$.
b)~Averaged number of bonds $N_B$ in steady state. Only model IV shows self-stabilization where the 
mean number of bonds increases initially with force. 
c)~Continuous rupture-rebinding transitions lead to a relative motion of the two planes bounding 
the adhesion. Self-stabilization reduces the motion.
d)~Relative deviation of the average number of unbound molecules $N_A$ from equilibrium. Note the 
increased molecule accumulation in the $a$ states for self-stabilization. See Supplementary 
Information I for parameters.
\label{fig:results:1}} 
\end{figure*}

\noindent 
\textbf{Self-stabilization of macromolecular adhesions.}~
We first consider the basic adhesion model, in which molecules do not change conformation, see 
Fig.~\ref{fig:intro}a and SI~movie~1.
Simulations reveal that a load $F$ can lead to a quasi-stationary adhesive state, where perpetual 
rupture and binding events result in a stationary state of persistent tangential sliding of the 
surfaces. For these adhesions, the mean number $\langle N_B \rangle$ of bonds always decreases monotonically with $F$, 
see Fig.~\ref{fig:intro}b. Therefore, increasing load on adhesions consisting of simple molecules 
promotes adhesion failure characterized by rupture of all bonds.
Next, we consider the generalized adhesion model consisting of molecules that can undergo an 
unfolding transition under force, see Fig.~\ref{fig:intro}c and SI~movie~2. For simplicity, we 
assume that unfolding only entails an increase in the rest length, while the elastic properties 
remain unchanged. Remarkably, the mean number $\langle N_B \rangle$ of bonds now initially grows 
with increasing load $F$, see Fig.~\ref{fig:intro}b, which prevents early adhesion failure. This 
striking effect, which we call ``self-stabilization'', is the central finding of this work. We 
want to emphasize that this effect crucially depends on the exchange of ``folded" molecules 
(state $a$) with the reservoir, so that the transitions $a \rightarrow b \rightarrow b_\mathrm{u}$ lead to a 
depletion of $a$ molecules, which can be replenished from the reservoir, thereby leading to an increase 
of the adhesion size. This mechanism is explained in detail below.

The simulation results can be corroborated with a mean-field approximation, which can be solved 
analytically \cite{Sabass2010}. The stationary distributions of molecules with extension $h$ in 
the bound and unbound states are denoted by $n_{b}(h)$ and $n_{a}(h)$, respectively. For the basic 
adhesion model without molecule unfolding, a drift-reaction equation is assumed where the average 
sliding velocity of the adhesion $v = \langle \dot{s} \rangle$ stretches the molecules, such that 
\begin{equation}
\partial_t n_b(h) + v \partial_h n_b(h) =\beta^+(h) n_a(h) - \beta^-(h) n_b(h)\,.
\label{eq:mf}
\end{equation}
Only stationary solutions with $\partial_t n_b(h)=0$ are considered. The total number of molecules 
in the adhesion is obtained as 
$N = N_B + N_A = \int_{-\infty}^{\infty}\![n_b(h)  +  n_a(h) ]\dif{h}$, where the extension of 
the unbound molecules obeys a Gaussian distribution, $n_a(h) \propto \mathcal{N}(0,\sigma_a^2)$. 
The non-linear equations are solved by expanding the distributions for small absolute values of 
$\tilde{v} = v /(k_\beta \sigma_b)$, where
the binding constant $k_\beta$ is employed as time unit and the extension variance of bound 
molecules $\sigma_b^2$ as length unit, e.g., 
$n_b(h) = n_b^*(h) + \tilde{v} n_{b_1}(h) + \frac{1}{2} \tilde{v}^2 n_{b_2}(h) + 
\mathcal{O}(\tilde{v}^3)$. The asterisk $(^*)$ here and in the following denotes equilibrium quantities 
calculated with $F=0$. Using the additional assumption that the optimal molecule extension for binding, 
$\ell_{\mathrm{b}}$, is much smaller than the typical length fluctuations, $\tilde{\ell}_\mathrm{b} = 
\ell_\mathrm{b} / \sigma_b \ll1$, we find
\begin{equation}
	N_{B} - N_{B}^* \approx -  (2/\pi)^{1/2} \tilde{\ell}_\mathrm{b} N_{B}^*\,\tilde{v}^2 \propto -F^2 \, ,
\end{equation}
and because, due to symmetry under reversal of the force direction, to leading order $\tilde{v}\propto\!F$. 
For the general case, where $\ell_{\mathrm{b}} \ll \sigma_b$ does not hold, the first non-vanishing 
correction to the equilibrium solution for the bonds $N_B$ can be shown to 
still be of the order $\tilde{v}^2$ and strictly negative, see Supplementary Information III. 
Thus, tangential force reduces the number of bonds and thereby always destabilizes simple adhesions consisting of molecules that do not undergo conformation changes, as expected intuitively.

To support the effect of self-stabilization in the generalized model shown in Fig.~\ref{fig:intro}c 
with analytical theory, we supplement Eq.~(\ref{eq:mf}) by two additional equations for binding and 
unfolding transitions, see Supplementary Information IV. 
For $|\tilde{v}| \ll 1$, the overall number of bound molecules, $N_B = N_b + N_{b_\mathrm{u}}$, can be expanded as
\begin{equation}
	N_B - N_B^* \approx N_{B,1} \tilde{v} + N_{B,2}\,\tilde{v}^2/2 \propto |F|.
\end{equation}
The leading contribution is linear in $\tilde{v}$ with a {\em positive} coefficient $N_{B,1}$ 
(obtained from  numerical analysis), as required for self-stabilization!
The second-order coefficient $N_{B,2}$ can be positive or negative, see Supplementary Information Fig.~S3.
Hence, analytical models confirm the existence of a self-stabilization regime where the number of adhesion bonds initially increases with load.

\noindent \textbf{Mechanism of self-stabilization.}~
To obtain deeper insight into the necessary ingredients for self-stabilization, and to demonstrate the 
essential contribution of the reservoir, we compare several variants of the basic models, see 
Fig.~\ref{fig:results:1} --- such as a simple 
adhesion model with fixed molecule number (model I), a model comprising a molecule reservoir and 
therefore adhesions of variable size (model II), a model with a fixed number of molecules that can 
unfold under force (model III), and a model combining unfolding molecules with a molecule reservoir 
(model IV). Some results from models II and IV are also shown in 
Fig.~\ref{fig:intro}. For models I-III, the mean number of bonds $N_B$ decreases with force, see 
Fig.~\ref{fig:results:1}b. For model II with variable system size, even the number of molecules in the 
adhesion decreases with force, leading to an earlier adhesion failure on average. 
Thus, neither a variable adhesion size nor molecule unfolding alone result in self-stabilization. In 
model IV, which combines a variable adhesion size with molecule unfolding, both the total number of 
molecules $N$ and the number of bonds $N_B$
initially grow with increasing load on the adhesion, Figs.~\ref{fig:results:1}a,b. The increased number 
of bonds improves load sharing among the molecules. One consequence is a significant reduction of the 
sliding motion of the adhesion, Fig.~\ref{fig:results:1}c. 
The inset in Fig.~\ref{fig:results:1}b shows that the bound fractions of molecules, $N_B/N$, as a 
function of $F$ collapse onto a single master curve for all models (I-IV). Hence, self-stabilization 
results from force-induced growth of the adhesion and not from changes of the rupture properties of 
individual molecules. This is the key difference to established catch-bond models, where individual 
molecules exhibit an increase of bond lifetime within limited force regimes.

Figure~\ref{fig:results:1}d illustrates the underlying mechanism of self-stabilization. The tangential 
load $F$ causes a continuous molecular-state turnover with recurring stretch, unfolding, and rupture of 
molecules along the transitions $a \rightarrow b\rightarrow b_\mathrm{u} \rightarrow a_\mathrm{u}$. 
Overall, the load increases the occupation of state $a_\mathrm{u}$.  Meanwhile, the state $a$, representing 
unbound, folded molecules, is in contact with the reservoir and molecules are replenished here, which 
allows a concurrent increase of the overall molecule number. The recruitment of molecules from the reservoir 
crucially depends on the intermediate state $a_\mathrm{u}$ {\em not} to be equilibrated, see 
Suppl.~Figs.~S6,S8, and to rebind and contribute to load sharing again via $a_\mathrm{u} 
\rightarrow b_\mathrm{u}$. 
It is important to note that without the reservoir, self-stabilization cannot occur because it requires 
growth of the adhesion cluster. 
Generically, the principle behind self-stabilization is that molecular-state occupation statistics are 
driven out of equilibrium in a way that results in further influx of molecules from a reservoir. 

\begin{figure*}
    \centering
    \includegraphics[width=0.9\linewidth]{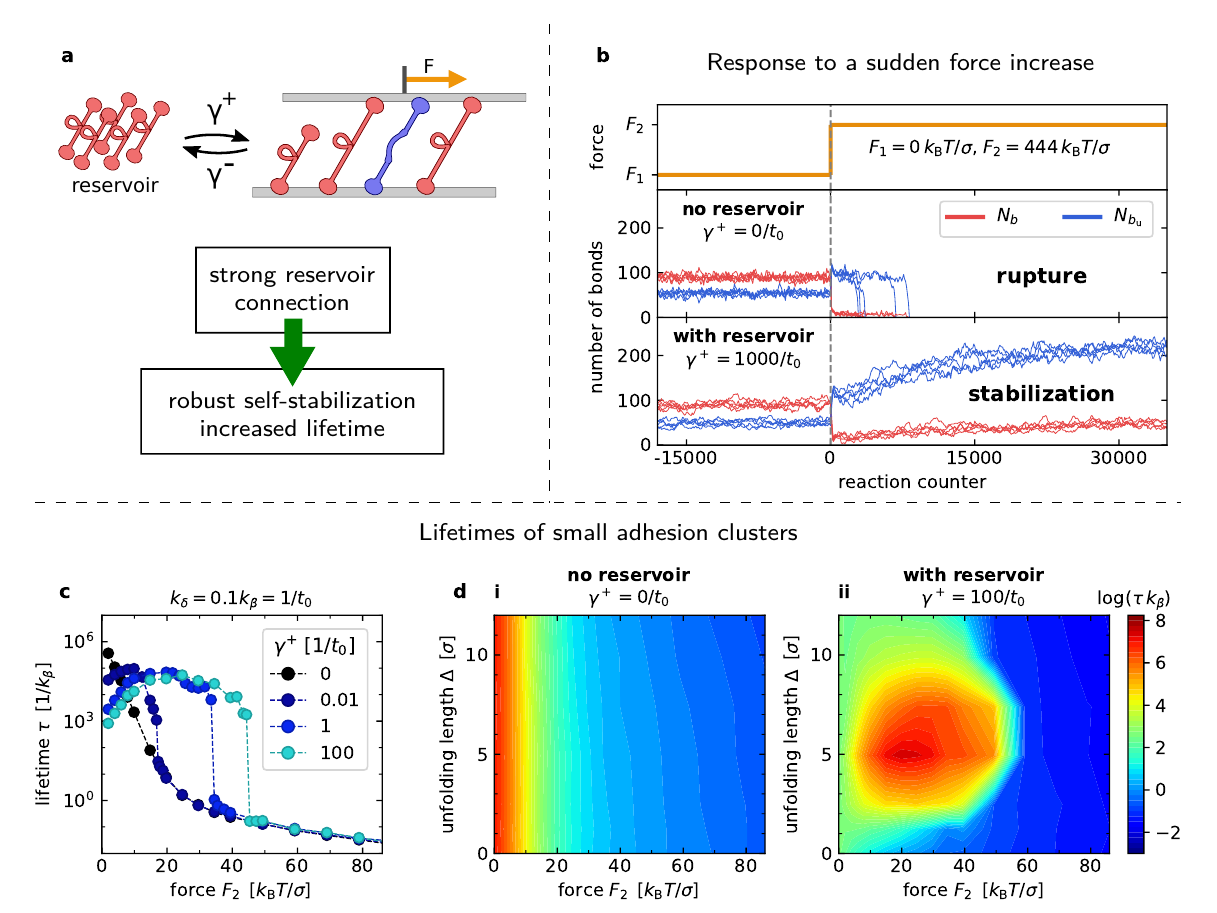}
    \caption{
Rupture behavior and lifetimes of adhesion clusters.
a)~The reservoir-exchange rates $\gamma^{\pm}$ control the association and dissociation of molecules with the adhesion. 
b)~Exemplary force-response of adhesion clusters without reservoir connection and with strong reservoir connection. A force jump amplifies molecule unfolding. Without self-stabilization, the cluster does not reach a non-equilibrium steady state but dissociates shortly after force application. 
c)~Average lifetimes of the adhesion clusters with $N^*\approx10$ for different values of $\gamma^\pm$ with $\gamma=\gamma^+/\gamma^-$ held constant.
A strong reservoir connection results in an adhesion lifetime maximum at finite, non-vanishing external forces.
d)~Average cluster lifetimes as a function of force and the unfolding length $\Delta$ for $k_\delta=0.1,\, k_\beta = 1/t_0$ and $\gamma^\pm = 0$ (i) and $\gamma^\pm=100/t_0$ (ii). See Supplementary I for other parameter values.
\label{fig:results:2}} 
\end{figure*}

\noindent \textbf{Self-stabilization is a robust mechanism.}~
To be an efficient and universally applicable mechanism, self-stabilization must be able to compensate load changes in non-stationary conditions and should not depend on a fine-tuning of 
parameters. To investigate these aspects, we consider step-like load changes for different model parameters. After a load jump from $F_1 = 0$ to $F_2 = F$, adhesion clusters either 
dissociate quickly or reach a new non-equilibrium steady state. Self-stabilization after a load 
jump depends on the strength of the molecule exchange with the reservoir, which is controlled 
by the values of $\gamma^+$ and $\gamma^-$, Fig.~\ref{fig:results:2}a.
Exemplary trajectories for the number of bonds in the folded and unfolded states, $b$ and $b_\mathrm{u}$, are shown in Fig.~\ref{fig:results:2}b.
The most-likely rupture forces are higher in self-stabilizing than in non-self-stabilizing adhesions, and grow 
with increasing reservoir-exchange rates ($\gamma^\pm > k_\beta$), see Supplementary Fig.~S4, because adhesion
molecules can quickly be recruited from or released into the reservoir.
The self-stabilization mechanism can thus also work under dynamic load conditions. 

To measure adhesion lifetimes, we simulate systems
consisting of few molecules, because lifetimes increase rapidly with the number of molecules. Load-jump simulations are carried out for a reservoir-exchange rate ratio 
$\gamma = 1$, which leads to adhesions with $N^{*}\approx10$ molecules in equilibrium. Lifetime is measured 
as the time from the force jump, $F_1 \rightarrow F_2$, to the rupture of the last adhesion bond. While lifetimes 
of adhesions with no reservoir coupling decrease monotonically with force, i.e., show a pure slip-bond behavior, 
lifetimes of self-stabilizing adhesions exhibit a maximum at non-vanishing forces,  see Fig.~\ref{fig:results:2}c. 
This lifetime maximum becomes more pronounced for increasing $\gamma^+$ and also depends on the rate-constant 
ratio $k_\delta/k_\beta$, see Supplementary Fig.~S5.
To further assess the robustness of the adhesion-lifetime increase to parameter choices, we vary the unfolding length $\Delta$, which determines the width of the energy barrier between the native and the unfolded molecule state. For adhesions without reservoir coupling, $\gamma^\pm=0$, the unfolding length $\Delta$ does not have a large impact on the adhesion lifetime, see Fig.~\ref{fig:results:2}d(i). 
However, a significant increase in adhesion lifetimes is observed for a broad range of unfolding lengths $\Delta>0$ if the reservoir coupling is strong, see Fig.~\ref{fig:results:2}d(ii). Self-stabilization is less effective for very small or very large values of $\Delta$, where native and unfolded state become indistinguishable or the unfolded state becomes inaccessible, respectively. 

\begin{figure*}
	\centering
	\includegraphics[width=\linewidth]{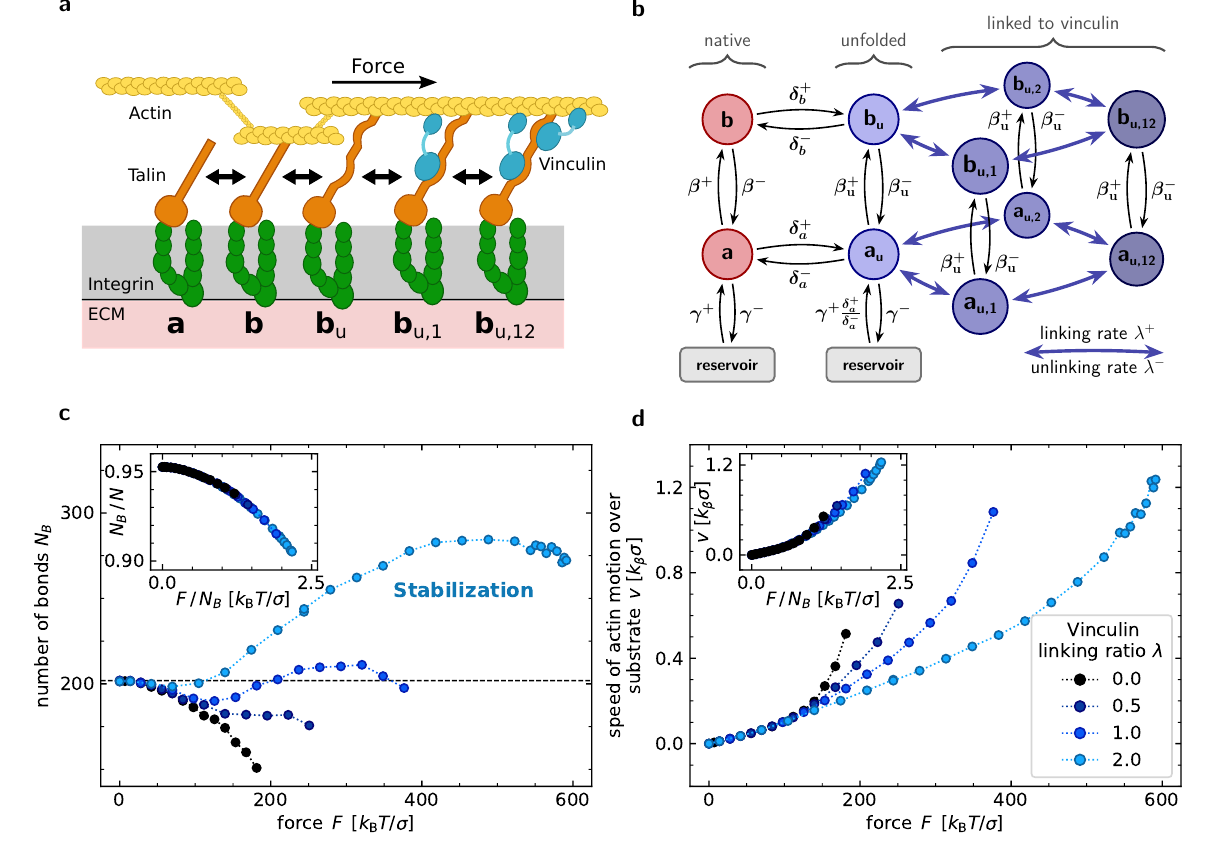}	
	\caption{
Exemplary realization of self-stabilization in focal adhesions. 
a)~Talin binds with its globular head to integrin and with its rod to actin filaments. The first domain to unfold under force is the R3 domain with two vinculin-binding sites. 
b)~State diagram for talin molecules.	
c-d)~Averaged steady-state simulation results for different linking ratios $\lambda = \lambda^+/\lambda^-$. See Supplementary I for parametrization.
c)~Vinculin recruitment produces self-stabilization.
d) The relative motion of the top plane modeling actin fibers is reduced through vinculin-based self-stabilization. 
\label{fig:results:talin}} 
\end{figure*}

\noindent 
\textbf{Cell-matrix adhesions.}~
The principle of self-stabilization can be realized in a large variety of molecular-state networks --- as long as they allow for non-equilibrated states that are increasingly populated under mechanical stress.
As stated in the introduction, talin unfolding and its interaction with other focal-adhesion proteins have been identified as major contributors to adhesion formation. We study here a model of self-stabilization that incorporates the binding of an adaptor protein such as vinculin to talin, see ~Fig.~\ref{fig:results:talin}a. The talin rod domain 
contains 11 cryptic vinculin binding sites. Under load, subdomains of the rod successively unfold, thereby 
enabling vinculin recruitment which blocks talin refolding and promotes focal-adhesion 
growth \cite{Yao2014,Atherton2016}.  
Talin unfolding typically starts at forces around $\SI{5}{\pico\newton}$ \cite{DelRio2009,Yao2014,Austen2015} 
with the R3 domain, by which two vinculin binding sites are exposed. In our model, we focus on this first 
unfolding transition. Six additional vinculin-bound states $b_{\mathrm{u},}$ and $a_{\mathrm{u},}$ are introduced 
as both sites in the R3 domain can be occupied independently, see~Fig.~\ref{fig:results:talin}b. Rate constants 
for binding and unbinding of vinculin to unfolded talin are denoted by $\lambda^\pm$ and are assumed to be the 
same for all transitions. Other model parameter values are estimated according to experimental results \cite{Liu2015, Yao2016, Tapia-Rojo2020}, see Supplementary Information I.

In the vinculin-bound states, talin refolding is blocked. Thus, vinculin binding generates talin states that are not in direct 
contact with the reservoir and can therefore be driven out of equilibrium. Simulation results for 
different values of the parameter $\lambda= \lambda^+/\lambda^-$ are shown in 
Fig.~\ref{fig:results:talin}c. For $\lambda=0$, all unbound states are equilibrated with the 
reservoir and no self-stabilization occurs. For $\lambda > 1$, the number of bonds increases 
with force, which not only stabilizes the adhesion but also translates into a reduction of adhesion 
sliding, see Fig.~\ref{fig:results:talin}d. Given the high affinity with a dissociation constant in 
the range $\num{e-7}-\SI{e-8}{\molar}$ of unfolded talin for vinculin, we expect the vinculin binding 
constant to be larger than unity~\cite{Bass2002, Chen2006, Wang2019, Tapia-Rojo2020}.
Moreover, we conjecture that the remaining cryptic binding sites in talin that open at higher forces 
extend the demonstrated self-stabilization effect to larger loads. In summary, the minimal adhesion 
model can be applied to specific biological systems like the integrin adhesome, where vinculin 
binding to unfolded talin domains results in a reaction network containing molecular states that are populated by force application --- and thus leads to self-stabilization. Analytical results 
of a simplified, corresponding mean-field model is given in Supplementary Information VI.

\section{Discussion}
Our theoretical model reveals a strikingly simple mechanism for a 
counter-intuitive load-response of adhesions, in which tangential mechanical load can result in 
adhesion enhancement --- instead of the ubiquitous adhesion weakening and rupture. This 
self-stabilization relies on molecular-conformation state-networks that are driven out of 
equilibrium by a mechanical load. By shifting the state occupations, the load causes a net influx 
of adhesion molecules from a surrounding reservoir. Notably, this self-stabilization does not 
require extra chemical energy, but the non-equilibrium conditions produced by the load suffice.

The primary motivation for our theoretical work are experimental observations on integrin-based focal adhesions, which display size adaptation to the applied load in planar cell cultures~\cite{riveline2001focal, balaban2001force}.
Different focal adhesion stabilization mechanisms presumably act in parallel, including actin polymerization, transcription regulation, integrin activation, and conformation changes of the adaptor protein talin.
Contrasting to this complexity, we predict that adhesion self-stabilization emerges naturally in 
systems that merely incorporate the unfolding transition of the adhesion molecules, like talin, 
and a mechanism which prevents rapid bulk-exchange of unfolded states, e.g., due to vinculin 
binding. Since the proposed models respect basic physical constraints, such as the 
detailed balance in thermal equilibrium, no energy is artificially injected into the system. 
Additional chemical driving, e.g., through the Rap1-GTP–interacting adaptor 
molecule~\cite{goult2013riam} or phosphorylation of vinculin or paxillin \cite{Zaidel-Bar2007, Stutchbury2017}, can provide an additional layer of biological control over the suggested adhesion-stabilization mechanism.

Mechanosensitive conformation changes of adhesion-linked proteins and subsequent recruitment of additional molecules are recurring motifs in many fundamental adhesion structures besides focal adhesions, for instance adherens junctions~\cite{yonemura2010alpha, le2010vinculin,Yao2014} and hemidesmosomes~\cite{zhang2011tension}. Motor proteins also undergo mechanosensitive conformation changes and can form dynamical ensembles. Therefore, we expect that the suggested non-equilibrium mechanism for self-stabilization can help to decipher many physiological and pathophysiological processes controlled by mechano-chemical factors, and may even allow novel designs of 
bio-inspired, artificial adhesion systems.

\section{Model and Methods}
\label{methods:section}
\subsection{Binding}
\label{methods:binding}
Different states are allowed to have their own spring constants and corresponding quantities are denoted by subscripts, e.g., $\kappa_a$ and $\sigma_a = \sqrt{k_\mathrm{B}T /\kappa_a}$.
The probability per unit time for an unbound molecule to bind at an extension $h$ is given by the function
\begin{equation}
    \beta^+(h) \frac{n_a(h)}{N_a} =   \frac{k_\beta}{\sqrt{2\pi} \sigma_b} \mathrm{e}^{\textstyle -\frac{(|h| - \ell_\mathrm{b})^2}{2 \sigma_b^2} + \frac{\epsilon_\mathrm{b}}{k_\mathrm{B}T}}
    \, ,
\end{equation}
where $k_\beta$ is the intrinsic binding rate constant and $\epsilon_\mathrm{b}$ is a constant. Similar binding rate expressions have been used previously~\cite{Bihr2012}. The total binding rate is obtained by integration over $h$ so that 
\begin{equation}
	\beta^+ = 
	k_\beta \left(1 + 
	\mathrm{Erf}\left(\tilde{\ell}_\mathrm{b}/\sqrt{2}\right) \right) 
	\mathrm{e}^{\tilde{\epsilon}_\mathrm{b} }
	\, ,
\end{equation}
where $\mathrm{Erf}(x)$ denotes the error function and the dimensionless quantities $\tilde{\ell}_\mathrm{b} = \ell_\mathrm{b}/\sigma_b$ and $\tilde{\epsilon}_\mathrm{b} = \epsilon_\mathrm{b}/(k_\mathrm{B}T)$ are used. Binding of unfolded molecules via $\beta_\mathrm{u}^+(h)$ is defined analogously with the extension $h_\mathrm{u}$ of unfolded molecules.

\subsection{Unbinding}
\label{methods:unbinding}
The rupture rate $\beta^-(h)$ is defined by the detailed-balance condition in thermal equilibrium~\cite{Dembo1988},
\begin{equation}
	\beta^+ (h)/\beta^- (h)  = 
	\mathrm{e}^{\textstyle -h^2/(2 \sigma_b^2) + h^2/(2\sigma_a^2) + \overline{\epsilon}_\mathrm{b} }
	\, ,
\end{equation}
where $\overline{\epsilon}_\mathrm{b} = \tilde{\epsilon}_\mathrm{b} + \ln(\sigma_a/\sigma_b)$ is the effective binding affinity.
The rupture rate results as
\begin{equation}
    \beta^-(h) = k_\beta 
    \mathrm{e}^{\textstyle (2 |h| \ell_\mathrm{b} - \ell_\mathrm{b}^2)/(2 \sigma_b^2) }
    \, .
\end{equation}
The rupture rate of unfolded bonds, $\beta_\mathrm{u}^-(h)$, is defined analogously with the extension $h_\mathrm{u}$ of unfolded molecules.

\subsection{Unfolding and refolding}
\label{methods:unfolding}
The unfolding and refolding reaction is modeled as the transition between two local energy minima separated by a single barrier. The distance to the barrier is denoted by $\Delta_1$ for unfolding and by $\Delta_2$ for refolding.
Their sum is equal to the total unfolding length $\Delta$. 
The unfolding rates are thus defined as
\begin{equation}
    \delta_{a,b}^+(h) = k_\delta 
    \mathrm{e}^{ \textstyle (2 \Delta_1 h - \Delta_1^2)/(2 \sigma_{a,b}^2)  - \tilde{\epsilon}_\mathrm{f} }
    \, ,
\end{equation}
where $\tilde{\epsilon}_\mathrm{f} = \epsilon_\mathrm{f}/(k_\mathrm{B}T)$ is a constant energy contribution for the conformation change.
The reverse rates are given by
\begin{equation}
     \delta_{a,b}^-(h) = k_\delta 
     \mathrm{e}^{\textstyle (-2 \Delta_2 h - \Delta_2^2)/(2 \sigma_{a,b}^2) }
     \, .
\end{equation}
The ratio of unfolding to refolding rate of bonds is given by the energy change of a bond going from extension $h$ to $h - \Delta$ as  
\begin{equation}
	\delta_b^+(h)/\delta_b^-(h - \Delta) = 
	\mathrm{e}^{\textstyle	(2h\Delta - \Delta^2)/(2\sigma_b^2)   
		- \tilde{\epsilon}_\mathrm{f}
	} \, .
\end{equation}
For unbound molecules, the total unfolding probability per time and bond is given by $\delta_a^+ = k_\delta \exp\left(-\tilde{\epsilon}_\mathrm{f} \right)\,$.
The total refolding rate is $\delta_a^- = k_\delta\,$.

\begin{acknowledgments}
BS acknowledges funding by the European Research Council (g.a.No. 852585)
\end{acknowledgments}

\section*{Author Contributions}

A.B. and B.S. designed and performed the research;
A.B., A.S., G.G. and B.S discussed results and wrote the manuscript.

\section*{Competing Interests statement}
The authors declare no competing interests.

\clearpage

\renewcommand{\thefigure}{S\arabic{figure}}
\renewcommand{\thetable}{S\arabic{figure}}
\setcounter{figure}{1}
\setcounter{table}{1}

\newpage
\setcounter{section}{0}
\renewcommand{\thesection}{Supplementary Discussion \Roman{section}} 
\renewcommand{\thesubsection}{\Roman{section}.\roman{subsection}}
\begin{widetext}
\section{Parameters}
\label{appendix:parameters}%
The minimal model for self-stabilization presented in the main text is motivated by focal cell-matrix adhesions. Model parameters were chosen accordingly.
Table~\ref{tab:appendix:parameter} contains all model parameters and their numerical values employed for the simulations if not stated otherwise in the text.
\\
\begin{table}[ht]
\centering
\renewcommand*{\arraystretch}{1.2}
\begin{tabular}{>{\centering}m{0.075\linewidth} m{0.3\linewidth}  >{\centering}m{0.175\linewidth}>{\centering}m{0.225\linewidth} >{\centering}m{0.075\linewidth} m{0.001\linewidth}}
	Variable	& Description	& sim. model	& sim. talin 	& unit &  \\  \hline
	$k_\mathrm{B}T$	& thermal energy 	&	
		\num{4.114}	& \num{4.114}	& \si{\pico\newton \nano\metre} & \\
	$\kappa_a$	& spring constant 	& 	
		\num{0.25}	& \num{0.5} 	&  \si{\pico\newton \per \nano\metre}&  \\
	$\kappa_b$	& spring constant 	& 	
		\num{0.25}	& \num{0.5} 	&  \si{\pico\newton \per \nano\metre}&  \\
	$\sigma$	& thermal fluctuation length	&	
		\num{4.057}	& \num{2.868} 	& \si{\nano\metre} & \\ \hline
	$k_\beta$		& rate, binding		&	
		\num{1} 	& \num{1}	& $1/t_0$ &\\
	$\ell_\mathrm{b}$		& binding distance	& 	
		\num{1} 	& \num{1}	&  \si{\nano\metre} & \\
	$\epsilon_\mathrm{b}$	& energy, binding	&	
		\num{1.50} 	& \num{3.00}	&  $k_\mathrm{B}T$ & \\
	$k_\delta$		& rate, folding	&  	
		\num{1} 	& \num{100}	&  $1/t_0$ & \\
	$\epsilon_\mathrm{f}$	& energy, folding	&	
		\num{0.50} 	& \num{5.83}	& $k_\mathrm{B}T$ & \\
	$\Delta	$	& unfolding length	&	
		\num{10}	& \num{12}	& \si{\nano\metre} & \\
	$\Delta_1$	& transition state distance, unfolding 	& 
		\num{5} 	& \num{7}	 & \si{\nano\metre} &  \\
	$\Delta_2$	& transition state distance, refolding	& 
		\num{5} 	& \num{5} 	& \si{\nano\metre} &  \\
	$\lambda^+$ 	& rate, linking		&	
		$\{0,1,2 \}$	&$\{0.0, 0.5, 1.0, 2.0 \}$	&  $1/t_0$ & \\
	$\lambda^-$ 	& rate, unlinking 	&	
		\num{1} 	& \num{1}	& $1/t_0$  & \\
	$\gamma^+$	& rate, addition from reservoir 	& 
		$\{20.0,14.5,11.4 \}$& $\{1., 0.996, 0.991, 0.977\}$	 & $1/t_0$  &	\\
	$\gamma^-$	& rate, removal to reservoir	& 
		\num{1} & 	\num{0.1}	 & $1/t_0$  & \\
\end{tabular}
\caption{Model parameters and values employed for simulations.
\label{tab:appendix:parameter}}
\end{table}

\noindent
In focal adhesions, mechanical force is transmitted along single or multiple proteins linked in series. In the latter case, the overall spring constant $\kappa_b$ will be dominated by the element with the smallest spring constant. 
The stiffness of cellular adhesion proteins lies in the order of $\si{\pico\newton\per\nano\metre}$ ~\cite{Kong2008,Roca-Cusachs2012}. The employed value of $\SI{0.25}{\pico\newton\per\nano\metre}$ is similar to values used for previous adhesion models~\cite{Erdmann2006,Schwarz2006,Qian2010,Fenz2017}.
The values of $\kappa_{a,b}$ used  for the talin simulations were chosen to match the experimentally measured unfolding and refolding behavior, as explained below. 
The thermal energy scale $k_{\mathrm{B}}T$ and the spring constant $\kappa$ determine the mean thermal fluctuation length $\sigma$, which is used as a length unit. 
\\
Experimentally, turnover of talin is significantly faster than turnover of integrin~\cite{gupton2006spatiotemporal}.
The binding and unbinding rates of talin are determined by the intrinsic rate $k_\beta$, the optimal binding distance $\ell_\mathrm{b}$, and the constant energy contribution $\epsilon_\mathrm{b}$. 
The intrinsic binding rate constant $k_\beta$ sets the time unit $t_0$. The value of the binding distance $\ell_\mathrm{b}$ agrees with the value employed in Ref.~\cite{Qian2010} and is smaller than the fluctuation length. The strength of individual bonds is determined by $F_0 = k_\mathrm{B}T/\ell_\mathrm{b} \approx \SI{4}{\pico\newton}$, as the rupture rate depends on force $F$ as $\propto\exp(F/F_0)$.
The constant $\epsilon_\mathrm{b}$ can be understood as an effective affinity parameter. Its value is chosen to be rather low to fix the adhesion cluster size at equilibrium, which is proportional to $\exp(\epsilon_\mathrm{b}/k_{\mathrm{B}}T)$, compare Refs.~\cite{Bihr2012, Bihr2015}. 
\\
The unfolding and refolding rates depend on $k_\delta$, $\epsilon_\mathrm{f}$, $\Delta$ and $\Delta_{1,2}$ with $\Delta = \Delta_1 + \Delta_2$. 
The energy contribution $\epsilon_\mathrm{f}$ determines the ratio between folded and unfolded molecules in equilibrium. It shifts the energy barrier between both states that are separated by the distances $\Delta_1$ and $\Delta_2$. 
For the simulations of talin molecules, the unfolding and refolding rate constants are chosen in accordance with the experimental results in Ref.~\cite{Yao2016}. 
For a time unit $t_0 = \SI{1}{\second}$, unfolding occurs at a rate of $\SI{0.015}{\per\second}$ at zero force. 
The folding and unfolding rates intersect at a force of $\SI{5}{\pico\newton}$, the value at which talin R3 domain unfolding and refolding is observed \cite{Yao2014,Yao2016,Tapia-Rojo2020}. 
\\
The cross-linking by additional adaptor proteins is described by making use of two rate constants $\lambda^\pm$. 
In the general simulation model, the value of these constants lies within the same order of magnitude as the intrinsic rate constants $k_\delta$ and $k_\beta$. 
For integrin-based adhesions, this reaction is realized by the association of vinculin to unfolded talin domains. 
The molecular interactions of talin and vinculin are highly complex. At low forces, vinculin binding at unfolded talin domains strengthens the adhesion \cite{Yao2014, Hirata2014, Ciobanasu2014}. 
The recruitment of vinculin is proposed to act as a negative feedback loop that stabilizes the force acting on the complex \cite{Tapia-Rojo2020}. 
Additionally, the vinculin-talin interaction also depends on the direction of forces~\cite{Kluger2020}.
In our model, we assume linking-rate constants with a value comparable to the intrinsic binding rate $k_\beta$.  
In recent experiments addressing the association of vinculin with the talin R3 domain, vinculin binding has been observed with a rate on the order of $\SI{e-1}{\per\second}$ when sufficient tension was applied to talin to induce unfolding~\cite{Tapia-Rojo2020}. 
\\
The exchange of molecules with a reservoir is governed by the two rate constants $\gamma^+$ and $\gamma^-$. Their ratio $\gamma$ is proportional to the number of molecules in the adhesion-cluster at equilibrium, i.e., when $F=0$. For comparability of the results of different models, the rate for adding molecules was increased from $\gamma^+$ to $\tilde{\gamma}^+ = \gamma^+ (1 + \exp(-\epsilon_\mathrm{f} / k_\mathrm{B}T))$ for the model without unfolding, and similarly when linking is included. This choice ensures the same number of molecules and bonds at $F=0$. 
As described in the main text, the molecule-exchange with the reservoir plays a fundamental role for the self-stabilization mechanism. A strong connection to the reservoir allows the adhesion to grow with increasing force. 

The construction of molecular state models for talin in integrin-based adhesions poses a challenge due to the large number of states and due to the fact that addition and removal of an adhesion molecule requires several steps \textit{in vivo} \cite{Moreno-Layseca2019,Kechagia2019}. 
Furthermore, several pathways have been suggested for the recruitment of talin to integrin-based adhesions \cite{Klapholz2017}. 
Values for the halftime of fluorescence recovery after photobleaching (FRAP) experiments for talin in focal adhesions are on the order of seconds or tens of seconds, depending on the cell type and substrate stiffness~\cite{Lele2008,Stutchbury2017}. 
The exchange of single molecules can therefore be expected to occur on timescales of seconds or even faster.
Hence, the values for $\gamma^\pm$ are chosen such, that they ensure a frequent exchange of molecules with the reservoir, but still allow to observe the processes within the cluster.

\section{Stochastic simulations}
\label{app:simulations}

Molecule-state trajectories are simulated with the Gillespie algorithm~\cite{Gillespie1976, Gillespie1977}.
Its basis is provided by the stochastic formulation of chemical kinetics where the probability that a reaction $i$ will happen in the infinitesimal time interval $\delta \tau$ is determined by the product of the available reactants or reactant pairs $N_i$ and a parameter $\mu_i$.
Assuming $j\in [1,\dots,m]$ available reactions, the function $P(i, \tau)$ describes the probability that the reaction $i$ is the first one to occur after a waiting time $\tau$ in the next infinitesimal time interval. For $0 \leq \tau < \infty$ we have
\begin{align}
P(i, \tau) = N_i \mu_i \exp\left(-\sum\limits_{j=1}^m N_j \mu_j \tau\right) \, .
\end{align}
Starting from an initial configuration, the time for the next reaction and the type of reaction are drawn according to $P(i,\tau)$ repeatedly. 
Force balance is restored instantaneously after each event.
\\
In response to the tangential external force, adhesion clusters either reach a quasi-steady-state or dissociate. The time- and ensemble averages are calculated for trajectories that do not lead to complete dissociation during the simulation time. If not stated otherwise, $50$ trajectories are tracked for $> \num{e6}$ single-bond transitions for the measurements of steady-state quantities, during which the tangential force is held constant. To measure adhesion lifetime, more than 200 cluster trajectories are simulated until their dissociation.

\section{Adhesion model without unfolding}
\label{appendix:expansion:nounfolding}
The following mean-field equations are used to approximate the dynamics of the stretch-dependent state-occupation numbers in the basic adhesion model without molecule unfolding
\begin{align}
	\diffp{}{t} N_a &= - \beta^+ N_a + \int\limits_{-\infty}^{\infty}\! \beta^-(h) n_b(h) \dif{}h  
     - \gamma^- N_a + \gamma^+ \, ,
     \label{eq_a_nounfolding}
     \\
    \diffp{}{t} n_b(h) &= - v \diffp{}{h} n_b(h) - \beta^-(h) n_b(h) + \beta^+(h) n_a(h)  \, ,
    \label{eq_b_nounfolding}
\end{align}
where we omitted explicit $t$-dependence in our notation.
Since the unbound molecules in state $a$ are assumed to relax quickly mechanically, their extensions obey a Gaussian distribution. Equation~(\ref{eq_a_nounfolding}) describes the mean total number of unbound molecules. 
The mean number of bonds $n_b(h)$ evolves according to Eq.~\eqref{eq_b_nounfolding}. The drift term in \eqref{eq_b_nounfolding} accounts for the average relative velocity $v = \langle \dot{s} \rangle$ between the adhesion planes that is due to the tangential force $F$. 
An expansion of $n_b(h)$ as
\begin{equation}
    n_b(h) = \sum\limits_{j=0}^{\infty} \frac{1}{j!}\, n_{b_j}(h)\, \tilde{v}^j,
\end{equation}
with $\tilde{v} = v/(k_\beta \sigma_b)$ solves the steady state conditions $\diffp{}{t} N_a = 0$ and $\diffp{}{t} n_b(h) = 0$ if
\begin{align}
    n_{b_0}(h) = n_b^*(h)
    = \frac{\gamma^+}{\gamma^-} \frac{\beta^+(h) }{\beta^-(h)} \, p_a(h)
        \, , \quad
	n_{b_j}(h) 
	= -j \frac{ n_{b_{j-1}}'(h) }{\beta^-(h)}  \quad \text{for}\ j>0 \, , 
\end{align}
where $p_a(h)$ is a Gaussian function with zero mean and variance $\sigma_a^2$ and the prime denotes the derivative with respect to the extension $h$. 
The first correction $n_{b_1}(h)$ is an odd function, so that its integral vanishes. 
The first non-vanishing correction is given by the term for $j=2$. 
Up to second order, the integrated steady state solution for the mean number of bonds $N_B = N_b$ is given by
\begin{align}
    N_B = \int\limits_{-\infty}^{\infty}\! n_b(h) \mathrm{d}h
    = N_B^*\left( 1  -  \sqrt{\frac{2}{\pi}} \tilde{\ell}_\mathrm{b} \mathrm{e}^{\tilde{\ell}_\mathrm{b}^2}\left( 1 - \sqrt{2\pi}\tilde{\ell}_\mathrm{b}\mathrm{e}^{2\tilde{\ell}_\mathrm{b}^2} \mathrm{Erfc}\left(\sqrt{2}\tilde{\ell}_\mathrm{b}\right)\right) \tilde{v}^2 \right) 
    \quad \mathrm{with}\ N_B^*=\frac{\gamma^+}{\gamma^-}\mathrm{e}^{\overline{\epsilon}_\mathrm{b}} \, ,
\end{align}
where $\tilde{v} = v/(\sigma_b k_\beta)$ and $\tilde{\ell}_\mathrm{b} = \ell_\mathrm{b}/\sigma_b$ have been used and $\mathrm{Erfc}(x)$ denotes the complementary error function. This first correction $\propto v^2$ is negative since the complementary error function for an argument $x > 0$ is bounded by
\begin{equation}
\mathrm{Erfc}(\sqrt{2} \ell_\mathrm{b}) \leq
    \frac{2}{\sqrt{\pi}} 
    \frac{\mathrm{e}^{-2\tilde{\ell}_\mathrm{b}^2}}{\sqrt{2} \tilde{\ell}_\mathrm{b} + \sqrt{2\tilde{\ell}_\mathrm{b}^2 + 4/\pi}}
    <
     \frac{\mathrm{e}^{-2\tilde{\ell}_\mathrm{b}^2}}{\sqrt{2\pi} \tilde{\ell}_\mathrm{b}}
     \, .
\end{equation}
The force-balance equation $F = \int_{-\infty}^{\infty} n_b(h) \kappa_b h\, \mathrm{d}h$ connects the external force with the resulting mean velocity $v$.
The lowest-order contribution ($j=0$) to the force balance vanishes because $n_{b_0}(h)$ is symmetric.
For the first non-vanishing correction, we find
\begin{equation}
    \tilde{F} = \frac{F \sigma_b}{ k_\mathrm{B}T} = 
    N_B^*
    \mathrm{e}^{\tilde{\ell}_\mathrm{b}^2/2}
    \left(
    -\frac{2 \tilde{\ell}_\mathrm{b}}{\sqrt{2\pi}} 
    + 
    \left(1 + \tilde{\ell}_\mathrm{b}^2\right) \mathrm{e}^{\tilde{\ell}_\mathrm{b}^2/2}
    \mathrm{Erfc}\left( 
        \frac{\tilde{\ell}_\mathrm{b}}{\sqrt{2}}
        \right)
    \right)
    \tilde{v}
    \, .
    \label{eq:app:firstcorrection1}
\end{equation}
The bracketed term on the right hand side of Eq.~(\ref{eq:app:firstcorrection1}) is strictly positive. The approximations for $\tilde{\ell}_\mathrm{b} \ll 1$ are given in the main text. 
\begin{figure}[htb]
\centering
\includegraphics[width=0.7\linewidth]{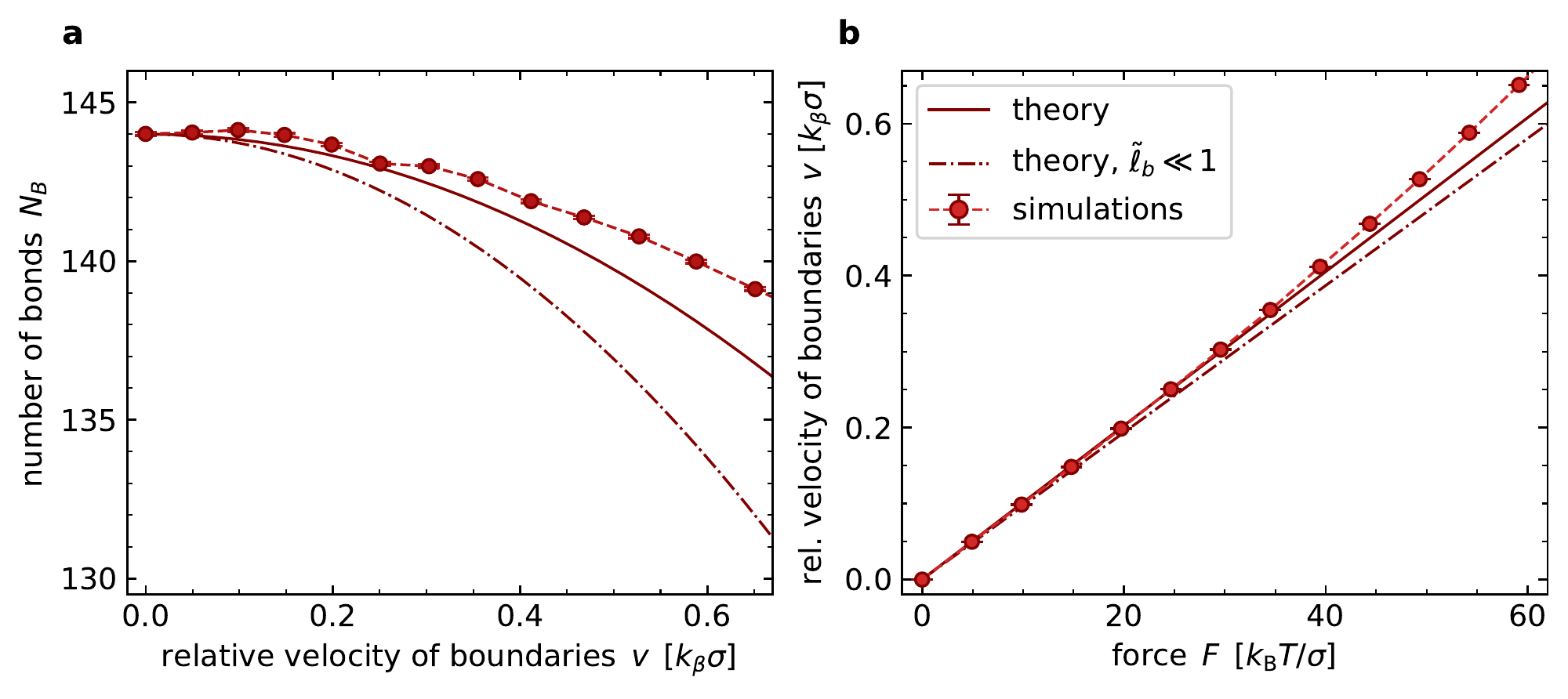}
\caption{
The average steady-state behavior of basic adhesions without unfolding. Symbols: Simulation results. Solid lines: analytical approximations using an expansion in $\tilde{v}$ up to second order. Dash-dotted lines without symbols: Approximation of the expansion for $\tilde{\ell}_\mathrm{b} \ll 1$. 
a)~The analytical approximation slightly underestimates the mean number of bonds in steady state. 
b)~The force-balance condition connects the external force $F$ with the drift $v$.  
\label{fig:nounfolding:theory}}
\end{figure}

\section{Adhesion model with unfolding}
\label{appendix:expansion:unfolding}
The numbers of unbound molecules obey the following approximate rate equations
\begin{align}
	\diffp{}{t} N_a &= - \beta^+ N_a  + \int\limits_{-\infty}^{\infty}\! \beta^-(h) n_b(h) \dif{}h  
	- \delta_a^+ N_a  + \delta_a^- N_{a_\mathrm{u}}  - \gamma^- N_a + \gamma^+ 
	\, ,
	\\
	\diffp{}{t} N_{a_\mathrm{u}} &= -  \beta_\mathrm{u}^+ N_{a_\mathrm{u}} + \int\limits_{-\infty}^{\infty}\! \beta_\mathrm{u}^-(h) n_{b_\mathrm{u}}(h) \dif{}h 
	 - \delta_a^- N_{a_\mathrm{u}}  +  \delta_a^+ N_a
	 \, .
\end{align}
The distributions of molecule numbers with extension $h$ experience a drift due to the velocity $v$ of the upper boundary and obey 
\begin{align}
	\diffp{}{t} n_b(h) &= - v \diffp{}{h} n_b(h) - \beta^-(h) n_b(h) + \beta^+(h) n_a(h) 
	 - \delta_b^+(h) n_b(h) + \delta_b^-(h) n_{b_\mathrm{u}}(h) 
	 \, ,
	 \\
	\diffp{}{t} n_{b_\mathrm{u}}(h) &= -v\diffp{}{h} n_{b_\mathrm{u}}(h) - \beta_\mathrm{u}^-(h) n_{b_\mathrm{u}}(h) + \beta_\mathrm{u}^+(h) n_{a_\mathrm{u}}(h) 
	- \delta_b^-(h) n_{b_\mathrm{u}}(h) + \delta_b^+(h) n_b(h) 
	\, .
\end{align}
The steady-state solution for $v=0$, i.e., the equilibrium solution, is given by
\begin{equation}
    N_a^* = \frac{\gamma^+}{\gamma^-} \, , \qquad
    N_{a_\mathrm{u}}^* = \frac{\gamma^+}{\gamma^-} \frac{\delta_a^+}{\delta_a^-} \, , \qquad
    n_b^*(h) = \frac{\gamma^+}{\gamma^-} \frac{\beta^+(h)}{\beta^-(h)}p_a(h) \, , \qquad
    n_{b_\mathrm{u}}^*(h) = \frac{\gamma^+}{\gamma^-} \frac{\beta_\mathrm{u}^+(h)}{\beta_\mathrm{u}^-(h)}p_{a_\mathrm{u}}(h) \, .
    \label{eq:steady_state_withunfolding}
\end{equation}
Integration yields the total number of bonds in equilibrium as
\begin{equation}
    N_B^* =  \int\limits_{-\infty}^{\infty}\! \Big[ n_b^*(h) + n_{b_\mathrm{u}}^*(h) \Big] \dif{}h
    = \frac{\gamma^+}{\gamma^-} 
    \mathrm{e}^{ \overline{\epsilon}_\mathrm{b} }
    \left(
        1 + \mathrm{e}^{-\tilde{\epsilon}_\mathrm{f}} 
    \right) 
    \, .
\end{equation}
To investigate the case $F>0$, the state distributions are expanded in powers of the average velocity $\tilde{v}$ as above, see \ref{appendix:expansion:nounfolding}. 
The connection of the state $a$ with the reservoir enforces $N_{a_j} = 0$ for $j>0$.
The solutions of the above equations cannot be found with an exact iterative formula. Instead, a numerical approach is used to find the corrections $N_{a_{\mathrm{u},j}}$, which can then be inserted back into the rate equations to solve for $n_{b_j}(h)$ and $n_{b_{\mathrm{u},j}}(h)$. Up to first order in $\tilde{v}$, the mean number of bonds remains unaffected or increases with the velocity, as demonstrated for different unfolding steps $\Delta$ and optimal binding lengths $\ell_\mathrm{b}$ in Fig.~\ref{fig:results:mathematica}a. 
The first-order correction vanishes for $\Delta \to 0$ and $\Delta \to \infty$. The second-order correction can be both positive or negative, Fig.~\ref{fig:results:mathematica}b. Strong self-stabilization is found for $\tilde{\ell}_\mathrm{b} \ll 1$. 
\begin{figure}[hbt]
\centering
\includegraphics[width=0.7\linewidth]{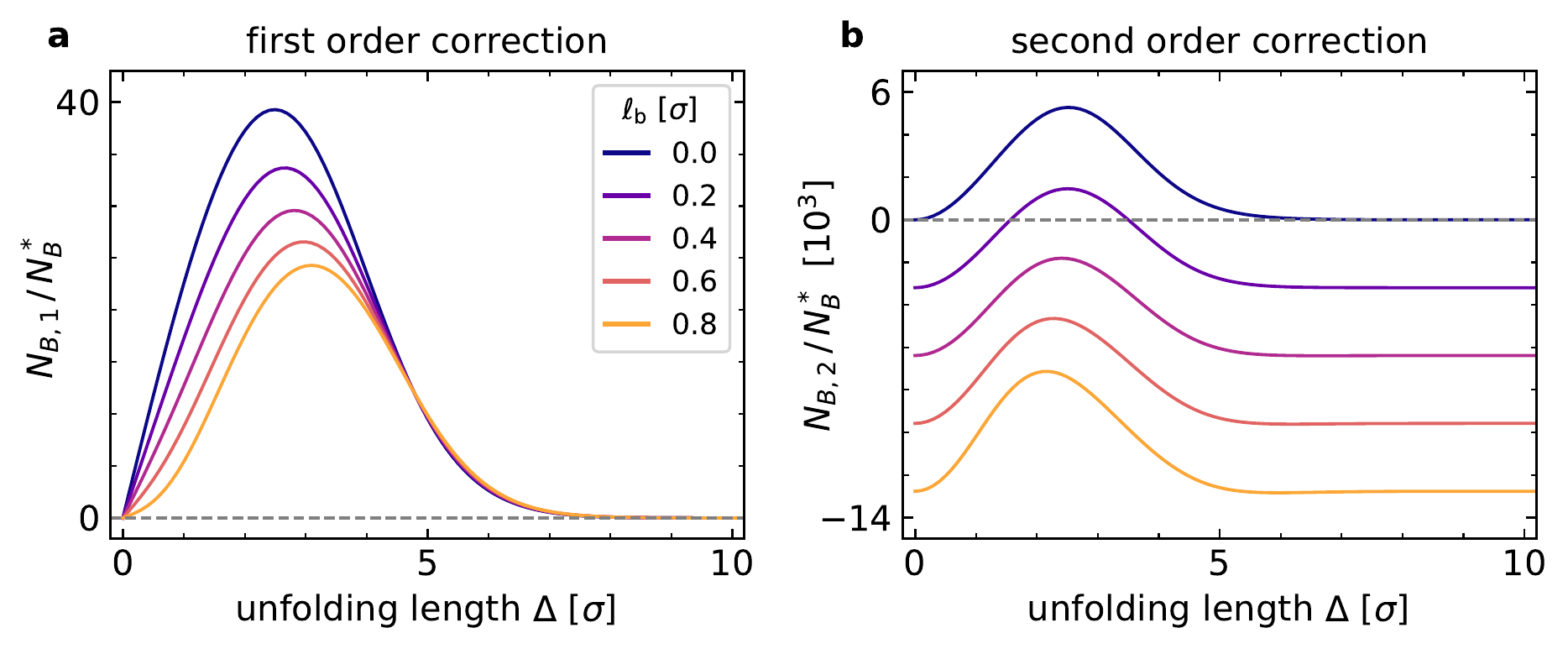}
\caption{
Numerical results for the coefficients of the first- and second-order corrections to the equilibrium solution for the number of bonds. The adhesion cluster connected to a molecule reservoir and results for different values for the molecule unfolding length $\Delta$ and the binding parameter $\ell_\mathrm{b}$ are shown. The first-order correction $\propto v$ is positive for $\ell_\mathrm{b} < \sigma_b$ and vanishes in the limits $\Delta \rightarrow 0$ and $\Delta \gg \sigma_b$.
The second-order correction $\propto v^2$ can be both positive and negative.
\label{fig:results:mathematica}} 
\end{figure}

\subsection{Rupture behavior and adhesion cluster lifetimes}
\noindent
Application of a sudden force jump from $F_1=0$ to $F_2>0$ either leads to a quick dissociation of the adhesion or the adhesion system relaxes to a non-equilibrium steady state. For large adhesion clusters ($N > 20$), the lifetimes of those adhesions that reach the non-equilibrium steady-state usually exceeds the finite simulation time. 
The fraction of those adhesion clusters that rupture almost immediately, $\Phi_\mathrm{rupt}$, is shown in Fig.~\ref{fig:app:rupturefraction}a as a function of $F_2$ for different reservoir rates $\gamma^+$. 
Without reservoir connection, when no self-stabilization occurs, the fraction of ruptured adhesion clusters sharply increases at a threshold force. Self-stabilization broadens the curves and increases the typical forces at which rupture occurs. The numerical results are well-fitted by a shifted and scaled error function.

The fraction of ruptured adhesion clusters after sufficiently long simulation times represents the cumulative probability distribution for cluster dissociation after a sudden force application. 
Therefore, the force value $F_2$ at which $\Phi_\mathrm{rupt}(F_2) = 0.5$ corresponds to the highest rupture probability, see Fig.~\ref{fig:app:rupturefraction}b. 
The most probable rupture force increases with increasing reservoir rate $\gamma^+$.
\begin{figure}[htbp]
\centering
\includegraphics[width=0.75\linewidth]{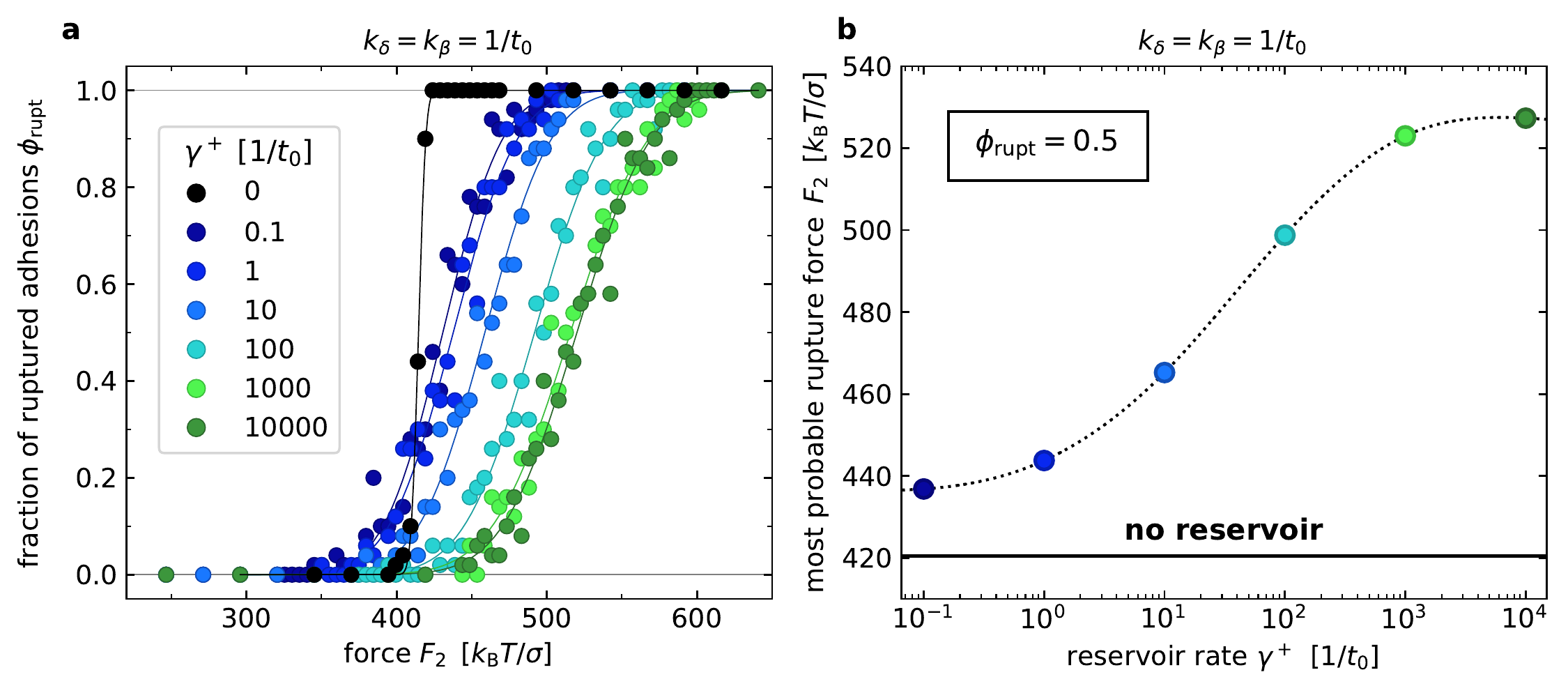}
\caption{
Fraction of ruptured adhesions after a sudden load-jump from $F_1=0$ to $F_2$
for different reservoir-exchange rates $\gamma^\pm$. The ratio $\gamma=\gamma^+/\gamma^- = 20$ is fixed. 
Systems were simulated for $\num{5e5}$ reaction steps after the force jump. 
a)~The fraction of ruptured adhesion clusters as a function of the magnitude of the force step. The strength of the reservoir connection, determined by $\gamma^\pm$, is varied while the ratio $\gamma = \gamma^+/\gamma^-$ is held constant. For comparison, the fraction of ruptured adhesion clusters in simulations without reservoir connection is shown with black bullets. A reservoir connection implies here self-stabilization.  
Due to the large system size ($N^* \approx 176$), most systems that did not rupture initially after the load jump remained stable for the rest of the simulation time. Lines show fits to error functions. 
b)~Force $F_2$ at which rupture is most likely for different reservoir rate values $\gamma^+$. Values are extracted from the fit curves in a). The black horizontal line shows the most probable rupture force for $\gamma^+=0$.
\label{fig:app:rupturefraction}}
\end{figure}
\begin{figure}[hbp]
\centering
\includegraphics[width=\linewidth]{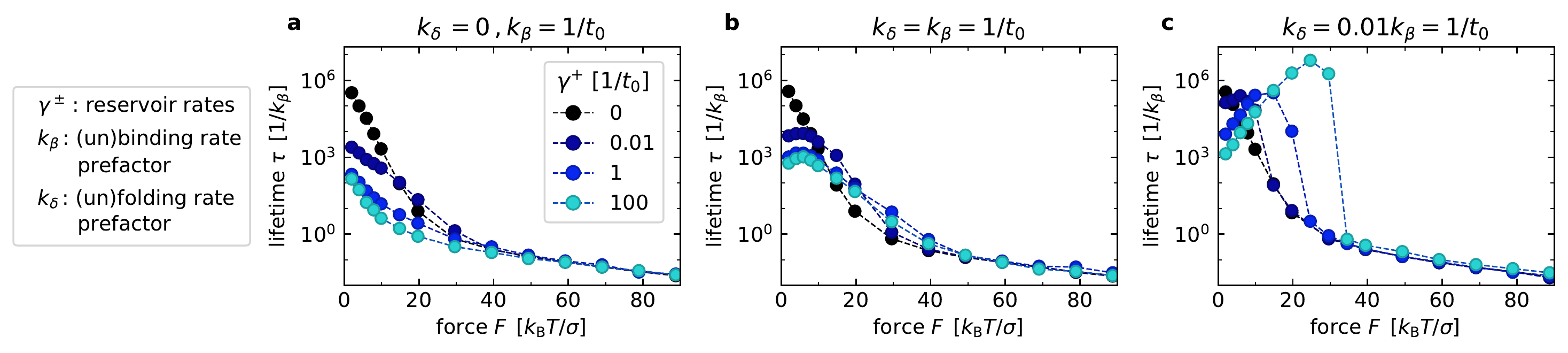}
\caption{
Lifetimes of small adhesions ($N^* \approx 10$) for different values of the rate prefactors $k_\beta$ and $k_\delta$ and the reservoir-exchange rates $\gamma^\pm$. The ratio $\gamma=\gamma^+/\gamma^- = 1$ is fixed. 
a)~$k_\delta$ = 0, no unfolding
b)~$k_\delta = k_\beta$  
c)~$k_\delta = 0.01 k_\beta$.
Only the lifetime results for the adhesion model IV, which exhibits self-stabilization, display a maximum at finite, non-vanishing forces. The lifetime of the adhesions is  substantially increased for intermediate forces if $\gamma^+$ and $k_\beta$ are larger than $k_\delta$, such that molecules coming from the reservoir establish new bonds frequently.
\label{fig:app:lifetimes}}
\end{figure}
\\
Fig.~\ref{fig:app:lifetimes} shows adhesion lifetimes, defined as the time until first complete dissociation of all bonds in an adhesion after application of a force jump. Small adhesion clusters ($N^* \approx 10$) with different values for the reservoir rates $\gamma^\pm$ are studied. The small molecule-numbers allow a direct measurement of the lifetimes by simulating the clusters until the last bond dissociates. For an adhesion system without molecule unfolding, realized by setting $k_\delta = 0$, the lifetime decreases monotonically with increasing force for any value of $\gamma^\pm$, see Fig.~\ref{fig:app:lifetimes}a. The equilibrium lifetime is largest for adhesion clusters without reservoir connection, realized by $\gamma^\pm = 0$. Figs~\ref{fig:app:lifetimes}b,c display lifetimes of adhesion models with unfolding molecules. The lifetime of adhesions without molecule exchange with a reservoir decrease monotonically with increasing force (black bullets). However, when the system is coupled to a molecule reservoir, so that the self-stabilization mechanism takes effect, the adhesion lifetime curves have a maximum at non-zero, finite forces. A substantial lifetime increase through self-stabilization is realized when both the reservoir rates and the intrinsic binding rate $k_\beta$ are large compared to the unfolding rate $k_\delta$. 

\subsection{Special case: no unfolding in the unbound state $a$}
\noindent
To investigate how cyclic fluxes along the state network $a - b - b_{\mathrm{u}} - a_{\mathrm{u}} - a$ affect self-stabilization, we set  $\delta_a^\pm = 0$. Thereby, the cycle in the single-molecule transition-diagram is broken. Physically, this modification means that unfolding is only allowed when the molecule is bound between both planes. It should be emphasized that this model variant still allows the emergence of cyclic fluxes in the high-dimensional continuous state space spanned by the extensions of the molecules.
\\
In Fig.~\ref{fig:app:deltaa0}, results from the new model variant are compared with results from model IV defined in the main text. For $\delta_a^\pm = 0$, the increase of the adhesion molecule number with force is significantly stronger than for the model IV with cyclic flux. Thus, self-stabilization is enhanced in the absence of cyclic flux, see Fig.~\ref{fig:app:deltaa0}a,b. 
The relative velocity of the two planes bounding the adhesion is reduced accordingly, see Fig.~\ref{fig:app:deltaa0}c. The enhanced self-stabilization can be attributed to a stronger accumulation of molecules in the state $a_\mathrm{u}$, from which molecules can only escape via state $b_{\mathrm{u}}$ in this model variant.
Note that the force value at which first rupture events are observed does not increase greatly, see black, vertical lines in Fig.~\ref{fig:app:deltaa0}a.  
A comparison with the inset in Fig.~\ref{fig:app:deltaa0}b shows, that the force per bond, $F/N_B$, is strongly reduced because of the increased number of bonds $N_B$. 
\begin{figure}[htbp]
\centering
\includegraphics[width=\linewidth]{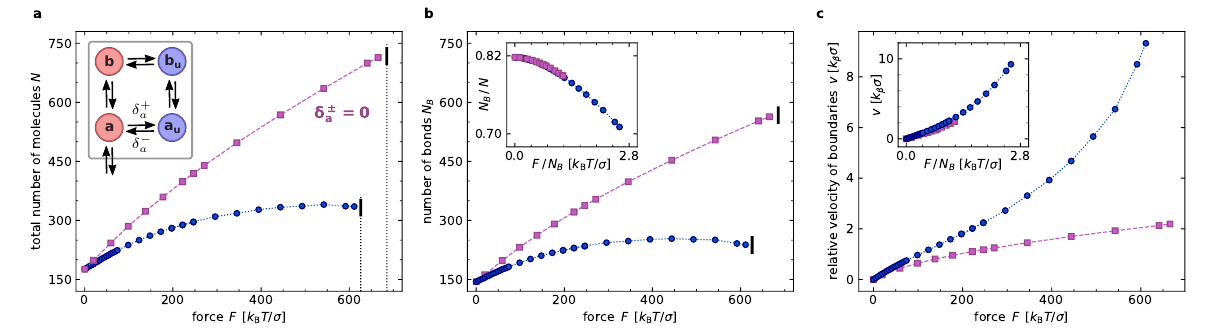}
\caption{
Steady state results of adhesion model IV (compare Fig.~2) and adhesion clusters with $\delta_a^\pm = 0$. When unbound unfolding and refolding is inhibited, the self-stabilization effect is more pronounced.
\label{fig:app:deltaa0}}
\end{figure}

\section{Adhesion model with unfolding and catch bonds}
\label{appendix:catchbonds}
In the models considered so far, a slip-bond behavior is assumed, i.e., the increase of the single-bond rupture rate with the applied extension $h$ is monotonic, see sec.~IV B.
Thus, so far, single-bond lifetimes decreased when tension increased.
Bonds that become longer-lived when tension increases are called catch bonds~\cite{Wang2019}. 
A number of molecular bonds in cellular adhesions have been described as catch-slip bonds ~\cite{Marshall2003,Kong2009,Chen2017, Manibog2014}. 
These interactions behave like a catch bond up to some force threshold while the slip-bond behavior takes effect at higher forces. To further test the generality of the self-stabilization mechanism, simulations of different adhesion models are performed where the bonds behave like pure catch bonds.
The binding and unbinding rate of folded bonds are changed to
\begin{equation}
    \beta^+(h)\frac{n_a(h)}{N_a} = \frac{k_\beta}{\sqrt{2\pi} \sigma_b} 
    \exp\left( -\frac{\left( |h| + \ell_\mathrm{b}\right)^2}{2\sigma_b^2} + \frac{\epsilon_\mathrm{b}}{k_\mathrm{B}T}\right)  
    \, , \qquad 
    \beta^-(h) = k_\beta\exp\left( \frac{-2 |h| \ell_\mathrm{b} - \ell_\mathrm{b}^2}{2\sigma_b^2} \right) \, .
\end{equation}
The rates for unfolded bonds are defined analogously with the extension $h_\mathrm{u}$. 
Note that the difference to the slip bond rates given in sec.~IV A-B lies only in one sign in each exponential function. 
Thus, detailed balance still holds in equilibrium at $F=0$.
The steady-state simulation results are shown in Fig.~\ref{fig:app:catchbonds}. 
Again, only for model IV with a combination of unfolding and association of new molecules from the reservoir, a pronounced increase in the number of bonds is observed for small forces, see Fig.~\ref{fig:app:catchbonds}b.
Figure~\ref{fig:app:catchbonds}d shows that the self-stabilization is caused by an accumulation of unbound molecules. 
Beyond this regime, at high forces, the pure catch-bond dynamics leads to a separation of the bond distribution into two subpopulations. Firstly, one has few, rather static molecules that carry most of the tension. Therefore, these molecules have a large extension and long lifetime. Secondly, a large number molecules form transient bonds with low, symmetrically distributed extensions.
As a result, the cluster stops moving, as can be seen in the velocity-force curve in Fig.~\ref{fig:app:catchbonds}c for high forces. 
\begin{figure}[hbt]
\centering
\includegraphics[width=\linewidth]{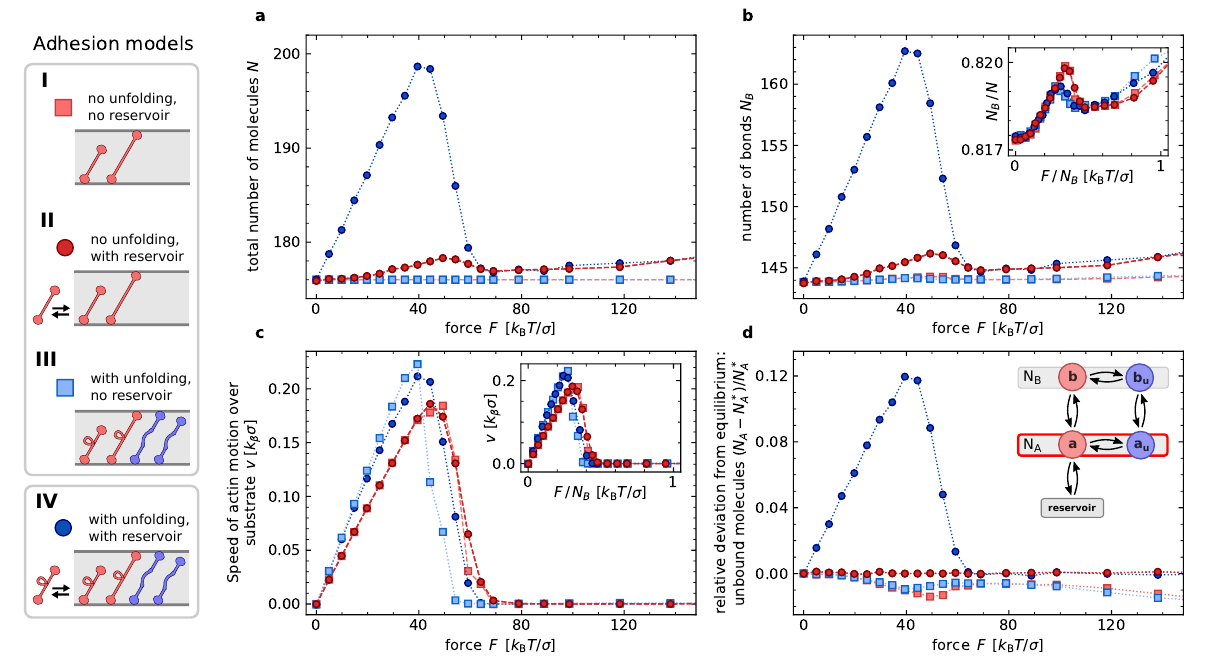}
\caption{
Simulation results for adhesion models in which single molecules behave like catch bonds. 
a)~Averaged total number of molecules $N$ in steady state.
b)~Averaged number of bonds $N_B$ in steady state.
Adhesion clusters with a reservoir connection (dark red and dark blue) show a regime, where the mean number of bonds increases with force. The self-stabilizing mechanism is particularly pronounced for the model IV (dark blue) in which molecules can unfold and are in exchange with a reservoir. 
c)~The continuous binding and rupture leads to an on average constant relative shift of the boundaries per time. Adhesion models with bond unfolding show an increased velocity.
d)~The relative deviation of the average number of unbound molecules $N_A$ from equilibrium shows that the self-stabilization mechanism is connected to an increased accumulation of adhesion molecules in unbound states (dark blue).
The parameter values are given in \ref{appendix:parameters}. 
\label{fig:app:catchbonds}}
\end{figure}

\section{Adhesion model with unfolding and cross-linking}
\label{appendix:expansion:linking}
We consider an extended adhesion model that also contains the effect of adaptor protein binding. Figure~\ref{fig:app:results:3}a shows the corresponding single-molecule state-transition diagram with linking rates $\lambda^{\pm}$. Linking the load-bearing molecules with adaptor proteins to the adhesion prevents a dissociation of the molecules from the adhesion. Thereby, the linking generates states that are not directly connected to the molecule reservoir.
\\
For an approximate analytical description of the system, the set of equations from \ref{appendix:expansion:unfolding} is extended to account for the linked states.
The unbound states obey
\begin{align}
	\diffp{}{t} N_a &= - \beta^+ N_a + \int\limits_{-\infty}^{\infty}\! \beta^-(h) n_b(h) \dif{}h  
	- \delta_a^+ N_a +  \delta_a^- N_{a_\mathrm{u}} 
	- \gamma^- N_a + \gamma^+ 
	\, ,
	\\
	\diffp{}{t} N_{a_\mathrm{u}} &= - \beta_\mathrm{u}^+ N_{a_\mathrm{u}} + \int\limits_{-\infty}^{\infty}\! \beta_\mathrm{u}^-(h) n_{b_\mathrm{u}} (h) \dif{}h 
	 -  \delta_a^- N_{a_\mathrm{u}} +  \delta_a^+ N_a
	 -  \gamma^- N_{a_\mathrm{u}} + \gamma^+ \frac{\delta_a^+}{\delta_a^-} 
	 -  \lambda^+ N_{a_\mathrm{u}} +  \lambda^- N_{a_{\mathrm{u},1}} 
	 \, ,
	 \\
	 	\diffp{}{t} N_{a_{\mathrm{u},1}} &= -  \beta_\mathrm{u}^+ N_{a_{\mathrm{u},1}}  + \int\limits_{-\infty}^{\infty}\! \beta_\mathrm{u}^-(h) n_{b_{\mathrm{u},1}}(h) \dif{}h 
	 -  \lambda^- N_{a_{\mathrm{u},1}}  +  \lambda^+ N_{a_{u}}
	 \, .
\end{align}

The equations determining the evolution of the extension-dependent state distributions read
\begin{align}
	\diffp{}{t} n_b(h) &= - v \diffp{}{h} n_b(h)  - \beta^-(h) n_b(h) + \beta^+(h) n_a(h) 
	 - \delta_b^+(h) n_b(h) + \delta_b^-(h) n_{b_\mathrm{u}}(h) 
	 \, ,
	 \\
	\diffp{}{t} n_{b_\mathrm{u}}(h) &= -v \diffp{}{h} n_{b_\mathrm{u}}(h)  - \beta_\mathrm{u}^-(h) n_{b_\mathrm{u}}(h) + \beta_\mathrm{u}^+(h) n_{a_\mathrm{u}}(h) 
	- \delta_b^-(h) n_{b_\mathrm{u}}(h) + \delta_b^+(h) n_b(h) 
	\nonumber 
	\\ &\quad 
    -  \lambda^+ n_{b_\mathrm{u}}(h) +  \lambda^- n_{b_{\mathrm{u},1}}(h)
    \, ,
    \\
    	\diffp{}{t} n_{b_{\mathrm{u},1}}(h) &= -v \diffp{}{h}   n_{b_{\mathrm{u},1}}(h)
    	- \beta_\mathrm{u}^-(h) n_{b_{\mathrm{u},1}}(h) + \beta_{u}^+(h) n_{a_{\mathrm{u},1}}(h) 
    -  \lambda^- n_{b_{\mathrm{u},1}}(h) +  \lambda^+ n_{b_{u}}(h)
    \, .
\end{align}
\begin{figure}[htb]
\centering
\includegraphics[width=0.85\linewidth]{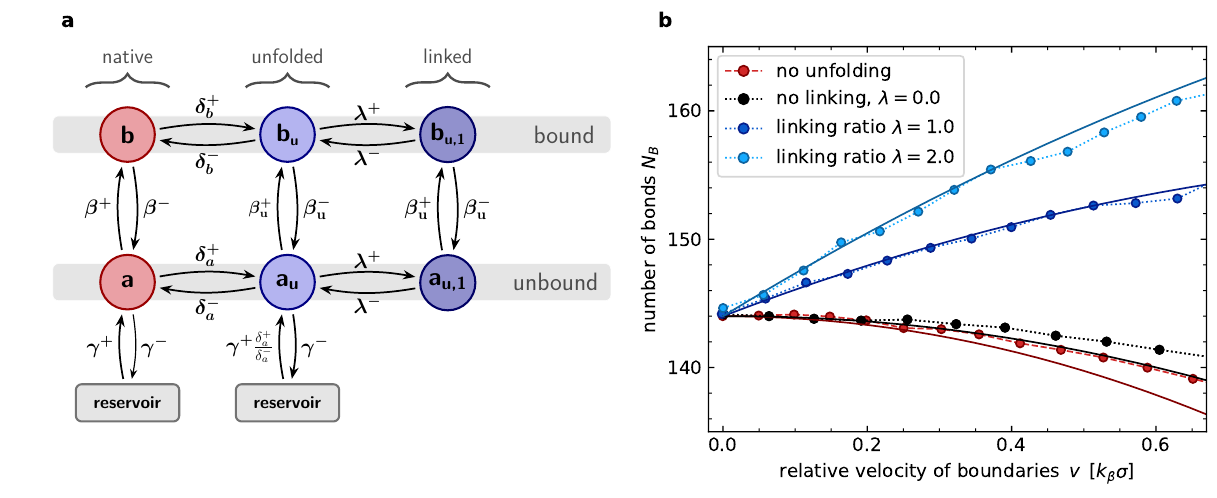}
\caption{
Model for adhesion clusters with self-stabilization realized by linking of auxiliary molecules. 
a)~Single-molecule state transition diagram. 
Unfolded molecules can establish an additional link to the adhesion that prevents one-step dissociation of the molecule from the adhesion.
b)~Comparison between simulation results (symbols connected by dashed lines) and the expansion of bond states up to second order in the speed $|\tilde{v}|$ (solid lines) for different linking ratios $\lambda$. 
Results for an adhesion model without the unfolding transition are shown for comparison. Parameter values are given in \ref{appendix:parameters}.
\label{fig:app:results:3}}
\end{figure}
\begin{figure}[hbp]
	\centering
	\includegraphics[width=0.85\linewidth]{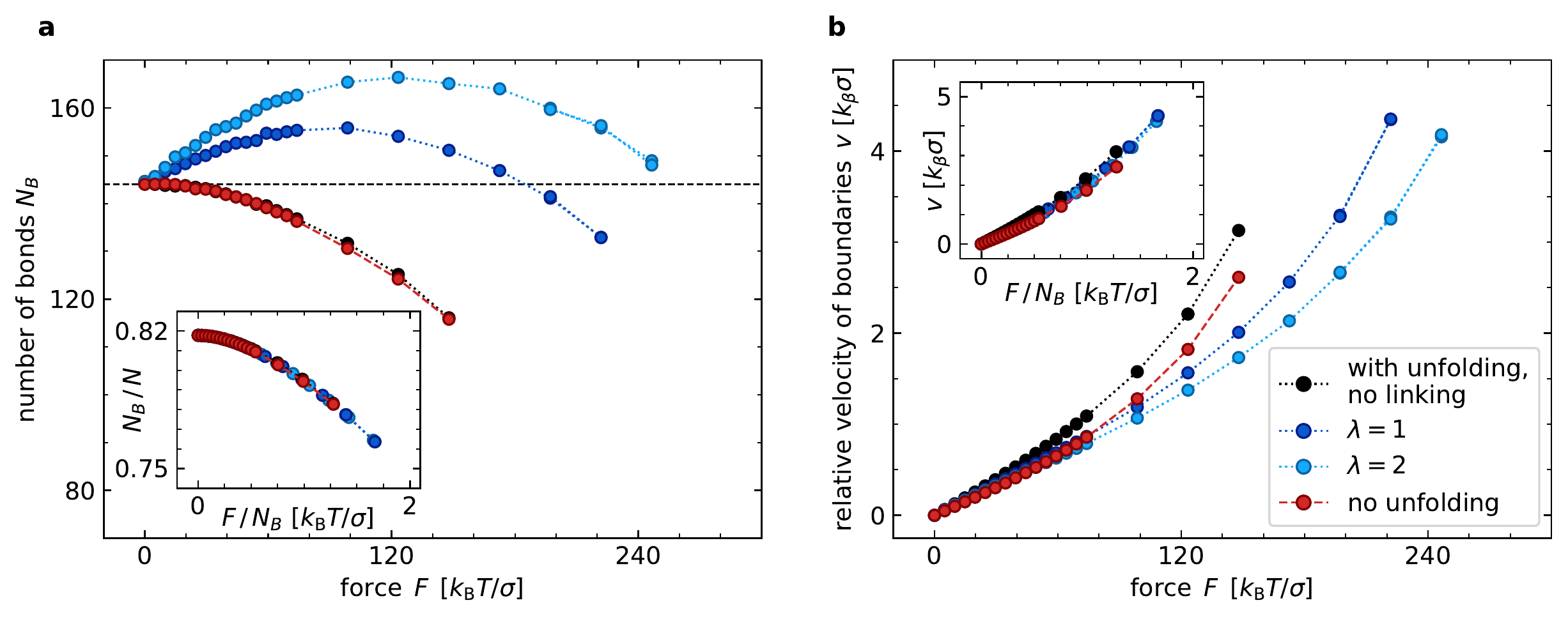}	
	\caption{
Averaged steady-state simulation results for adhesion models with the unfolding-linking mechanism and two reservoir connections.
Black and blue symbols show results for different linking ratios $\lambda = \lambda^+/\lambda^-$.
Red: Adhesions without unfolding for comparison. 
a)~Only adhesions with a positive linking ratio show a self-stabilization regime where the number of bonds increases with force.
b)~Relative motion of the upper plane and the adhesion. Insets: The fraction of bonds and the velocity as a function of force per bond show the same behavior for all systems.
The parameter values are given in \ref{appendix:parameters}.
\label{fig:app:results:2}} 
\end{figure}

The steady-state results for the number of molecules in state $a$, $b$, $b_\mathrm{u}$, $a_\mathrm{u}$ are given by Eq.~\ref{eq:steady_state_withunfolding}.
The additional linked states fulfill
\begin{equation}
    n_{b_{\mathrm{u},1}}^*(h) = \frac{\lambda+}{\lambda^-} n_{b_\mathrm{u}}^*(h) \, , \quad  
    N_{a_{\mathrm{u},1}}^* = \frac{\lambda+}{\lambda^-} N_{a_\mathrm{u}}^* 
    \, .
\end{equation}
As before, an expansion of the distributions for small speeds $|\tilde{v}|$ yields the first corrections to the equilibrium distribution. Figure~\ref{fig:app:results:3} shows a comparison between steady-state results obtained in simulations and in the analytical approximation. Figure~\ref{fig:app:results:2} shows simulation results for a range of parameters.

\clearpage

\end{widetext}

\bibliography{mainbib}

\end{document}